 \documentclass{emulateapj}

\newcommand{\Kepler}{{\em Kepler}}
\newcommand{\corot}{{\em CoRoT}}
\newcommand{\numax}{\mbox{$\nu_{\rm max}$}}
\newcommand{\Dnu}{\mbox{$\Delta \nu$}}
\newcommand{\dnu}[1]{\mbox{$\delta \nu_{#1}$}}
\newcommand{\muHz}{\mbox{$\mu$Hz}}
\newcommand{\half}{{\textstyle\frac{1}{2}}}

\newcommand{\new}[1]{{\bf #1}}
\renewcommand{\new}[1]{{#1}}

\slugcomment{Accepted by ApJ Letters}

\shorttitle{Oscillations in low-luminosity red giants with \Kepler}
\shortauthors{T. R. Bedding et al.}

\begin{document}

\title{Solar-like oscillations in low-luminosity red giants: first results
from \em Kepler}

\author{
T.~R.~Bedding,\altaffilmark{1} 
D.~Huber,\altaffilmark{1} 
D.~Stello,\altaffilmark{1} 
Y.~P.~Elsworth,\altaffilmark{2} 
S.~Hekker,\altaffilmark{2} 
T.~Kallinger,\altaffilmark{3,4} 
S.~Mathur,\altaffilmark{5} 
B.~Mosser,\altaffilmark{6} 
H.~L.~Preston,\altaffilmark{7,8} 
J.~Ballot,\altaffilmark{9} 
C.~Barban,\altaffilmark{6} 
A.~M.~Broomhall,\altaffilmark{2} 
D.~L.~Buzasi,\altaffilmark{7} 
W.~J.~Chaplin,\altaffilmark{2} 
R.~A.~Garc\'\i a,\altaffilmark{10} 
M.~Gruberbauer,\altaffilmark{11} 
S.~J.~Hale,\altaffilmark{2} 
J.~De~Ridder,\altaffilmark{12} 
S.~Frandsen,\altaffilmark{13} 
W.~J.~Borucki,\altaffilmark{14} 
T.~Brown,\altaffilmark{15} 
J.~Christensen-Dalsgaard,\altaffilmark{13} 
R.~L.~Gilliland,\altaffilmark{16} 
J.~M.~Jenkins,\altaffilmark{17} 
H.~Kjeldsen,\altaffilmark{13} 
D.~Koch,\altaffilmark{14} 
K.~Belkacem,\altaffilmark{18} 
L.~Bildsten,\altaffilmark{19} 
H.~Bruntt,\altaffilmark{6,13} 
T.~L.~Campante,\altaffilmark{13,20} 
S.~Deheuvels,\altaffilmark{6} 
A.~Derekas,\altaffilmark{1,23} 
M.-A.~Dupret,\altaffilmark{18} 
M.-J.~Goupil,\altaffilmark{6} 
A.~Hatzes,\altaffilmark{21} 
G.~Houdek,\altaffilmark{4} 
M.~J.~Ireland,\altaffilmark{1} 
C.~Jiang,\altaffilmark{22} 
C.~Karoff,\altaffilmark{2} 
L.~L.~Kiss,\altaffilmark{1,23} 
Y.~Lebreton,\altaffilmark{6} 
A.~Miglio,\altaffilmark{18} 
J.~Montalb\'an,\altaffilmark{18} 
A.~Noels,\altaffilmark{18} 
I.~W.~Roxburgh,\altaffilmark{24} 
V.~Sangaralingam,\altaffilmark{2} 
I.~R.~Stevens,\altaffilmark{2} 
M.~D.~Suran,\altaffilmark{25} 
N.~J.~Tarrant,\altaffilmark{2} and 
A.~Weiss\altaffilmark{26} 
}
\altaffiltext{1}{Sydney Institute for Astronomy (SIfA), School of Physics, University of Sydney, NSW 2006, Australia; \mbox{bedding@physics.usyd.edu.au}}
\altaffiltext{2}{School of Physics and Astronomy, University of Birmingham, Birmingham B15 2TT, UK}
\altaffiltext{3}{Department of Physics and Astronomy, University of British Columbia, Vancouver, Canada}
\altaffiltext{4}{Institute of Astronomy, University of Vienna, 1180 Vienna, Austria}
\altaffiltext{5}{Indian Institute of Astrophysics, Koramangala, Bangalore 560034, India}
\altaffiltext{6}{LESIA, CNRS, Universit\'e Pierre et Marie Curie, Universit\'e Denis, Diderot, Observatoire de Paris, 92195 Meudon cedex, France}
\altaffiltext{7}{Eureka Scientific, 2452 Delmer Street Suite 100, Oakland, CA 94602-3017, USA}
\altaffiltext{8}{Department of Mathematical Sciences, University of South Africa, Box 392 UNISA 0003, South Africa}
\altaffiltext{9}{Laboratoire d'Astrophysique de Toulouse-Tarbes Universit\'e de Toulouse, CNRS 14 av E. Belin 31400 Toulouse, France}
\altaffiltext{10}{Laboratoire AIM, CEA/DSM-CNRS, Universit\'e Paris 7 Diderot, IRFU/SAp, Centre de Saclay, 91191, Gif-sur-Yvette, France}
\altaffiltext{11}{Department of Astronomy and Physics, Saint Mary's University, Halifax, NS B3H 3C3, Canada}
\altaffiltext{12}{Instituut voor Sterrenkunde, K.U.Leuven, Belgium}
\altaffiltext{13}{Danish AsteroSeismology Centre (DASC), Department of Physics and Astronomy, Aarhus University, DK-8000 Aarhus C, Denmark}
\altaffiltext{14}{NASA Ames Research Center, MS 244-30, Moffett Field, CA 94035, USA}
\altaffiltext{15}{Las Cumbres Observatory Global Telescope, Goleta, CA 93117, USA}
\altaffiltext{16}{Space Telescope Science Institute, 3700 San Martin Drive, Baltimore, Maryland 21218, USA}
\altaffiltext{17}{SETI Institute/NASA Ames Research Center, MS 244-30, Moffett Field, CA 94035, USA}
\altaffiltext{18}{Institut d'Astrophysique et de G\'{e}ophysique de l'Universit\'{e} de Li\`{e}ge, All\'ee du 6 Ao\^{u}t 17 - B 4000 Li\`{e}ge, Belgium}
\altaffiltext{19}{Kavli Institute for Theoretical Physics and Department of Physics, University of California, Santa Barbara, CA 93106, USA}
\altaffiltext{20}{Centro de Astrof\'isica da Universidade do Porto, Rua das Estrelas, 4150-762 Porto, Portugal}
\altaffiltext{21}{Thueringer Landessternwarte Tautenburg, Sternwarte 5, D-07778, Tautenburg, Germany}
\altaffiltext{22}{Department of Astronomy, Beijing Normal University, China}
\altaffiltext{23}{Konkoly Observatory of the Hungarian Academy of Sciences, H-1525 Budapest, P.O. Box 67, Hungary}
\altaffiltext{24}{Queen Mary University of London, Mile End Road, London E1 4NS, UK}
\altaffiltext{25}{Astronomical Institute of the Romanian Academy, Str.\ Cutitul de Argint, 5, RO 40557,Bucharest, Romania}
\altaffiltext{26}{Max-Planck-Institut f\"ur Astrophysik, Karl-Schwarzschild-Str.~1, 85748 Garching, Germany}

\begin{abstract}
We have measured solar-like oscillations in red giants using time-series
photometry from the first 34 days of science operations of the {\em Kepler
Mission}.  The light curves, obtained with 30-minute sampling, reveal clear
oscillations in a large sample of G and K giants, extending in luminosity
from the red clump down to the bottom of the giant branch.  We confirm a
strong correlation between the large separation of the oscillations (\Dnu)
and the frequency of maximum power (\numax).  We focus on a sample of 50
low-luminosity stars ($\numax > 100\,\muHz$, \new{$L \lesssim 30\,L_\sun$})
having high signal-to-noise ratios and showing the unambiguous signature of
solar-like oscillations.  These are H-shell-burning stars, whose
oscillations should be valuable for testing models of stellar evolution and
for constraining the star-formation rate in the local disk.  We use a new
technique to compare stars on a single \'echelle diagram by scaling their
frequencies and find well-defined ridges corresponding to radial and
non-radial oscillations, including \new{clear evidence for} modes with
angular degree $l=3$.  Measuring the small separation between $l=0$ and
$l=2$ allows us to plot the so-called C-D diagram of \dnu{02}\ versus \Dnu.
The small separation \dnu{01} of $l=1$ from the midpoint of adjacent $l=0$
modes is negative, contrary to the Sun and solar-type stars.  The ridge for
$l=1$ is notably broadened, which we attribute to mixed modes, confirming
theoretical predictions for low-luminosity giants.  Overall, the results
demonstrate the tremendous potential of {\em Kepler} data for
asteroseismology of red giants.
\end{abstract}


\section{Introduction}

Studying solar-like oscillations is a powerful way to probe the interiors
of stars \citep[e.g.,][]{B+G94,ChD2004}.  Oscillations in main-sequence
stars and subgiants have been measured using ground-based spectroscopy
\citep[for a recent review see][]{AChDC2008} and more recently by space
missions such as \corot\ \citep[e.g.,][]{MBA2008}.  Red giants oscillate at
lower frequencies and so demand long and preferably uninterrupted time
series to resolve their oscillations.  

The first indications of solar-like oscillations in G and K giants were
based on ground-based observations in radial velocity (see \citealt{Mer99}
and references therein; \citealt{frandsen02,deridder06}) and photometry
\citep{stello06}, and on space-based photometry from
the {\em Hubble Space Telescope} ({\em HST}; \citealt{E+G96,gilliland08,stello09b}),
{\em WIRE} \citep{buzasi00,retter03,stello08}, 
{\em Microvariability and Oscillations of Stars} ({\em MOST};  \citealt{BMDeR2007,KGW2008,KGM2008}), and
{\em Solar Mass Ejection Imager} ({\em SMEI}; \citealt{tarrant07}).
A major breakthrough came from observations over about 150 days with the
\corot\ space telescope, which produced clear detections in numerous stars
of both radial and non-radial oscillations in the frequency range
10--100\,\muHz\ \citep{DeRBB2009,HKB2009,CDRB2010}.  This increased the
number of known pulsating G and K giants to nearly 800, and enabled the
first systematic study of oscillations in a large population of red clump
stars \citep{HKB2009,MMB2009,KWD2010}.  \new{Note that the red clump
comprises core He-burning stars and is the metal-rich counterpart to the
horizontal branch.}

Detecting oscillations in lower-luminosity red giants, which are
H-shell-burning stars at the base of the giant branch, is very desirable
for testing models of stellar evolution and also for constraining the
star-formation rate in the local disk \citep{MMB2009}.  However, such
detections are even more challenging than for clump stars because the
amplitudes are substantially lower.  In addition, the oscillation
frequencies are higher and are comparable to the orbital frequency of a
satellite in low-Earth orbit.  These factors made it difficult for \corot\
to achieve such detections.

In this Letter, we present data from the asteroseismology program of the
{\em NASA} \Kepler\ Mission \citep{GBChD2010} that reveal clear
oscillations in a large sample of red giants, extending in luminosity from
the red clump down to the bottom of the giant branch.

\section{Global Oscillation Parameters}

The observations were obtained over the first 34 days of science operations
of the \Kepler\ Mission (Q1 data).  We analyzed light curves having
29.4-minute sampling (long-cadence mode) for about 1500 stars listed as red
giants in the {\em Kepler Input Catalog} (KIC; \citealt{LBM2005}).  About
$20\%$ of the Q1 light curves showed large variations ($>$1\%) on long
timescales, mostly due to M-giant pulsations, and are not analyzed
here.\footnote{One of these was subsequently found to be an oscillating red
giant in an eclipsing binary system \citep{HDH2010}.}  For the remaining
sample, we used the light curves to measure the global oscillation
parameters, most notably the frequency of maximum power (\numax) and the
mean large frequency separation (\Dnu).  We were able to find a power
excess in $\sim$1000 stars \citep[see also Figure~5 in][]{GBChD2010} and
measure the large separation for $\sim$700 stars.  In this Letter we focus
on the low-luminosity red giants (corresponding to $\numax > 100\,\muHz$)
for which, as discussed in the introduction, \corot\ results have not yet
been reported.  As discussed by \citet{MMB2009}, such stars generally have
luminosities below about $30\,L_\sun$.

\begin{figure}
\epsscale{0.6}
\plotone{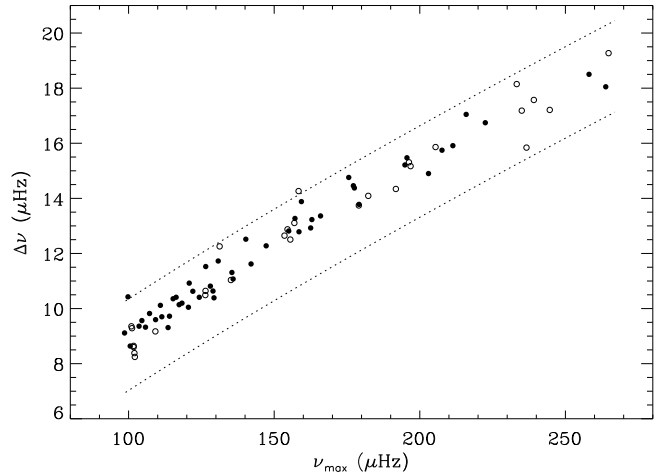}
\caption{Large frequency separation versus frequency of maximum power for
  78 low-luminosity red giants.  Filled symbols indicate the 50 stars that
  were selected for more detailed analysis. The dotted lines enclose the
  region containing $\sim$80\% of measurements by the five pipelines (see
  text). }
\label{fig:01}
\end{figure}

Figure~\ref{fig:01} shows the measured values of \Dnu\ and \numax\ for
these stars.  The symbols show results obtained using the pipeline
described by \citet{HSB09}, \new{who also described the method used to fit
and subtract the background.}  Similar results were found using pipelines
developed by four other groups \citep{HBC2010,M+A2009,KWD2010,MGR2010}, and
the dotted lines in Figure~\ref{fig:01} show the region inside which
$\sim$80\% of those measurements lay.  A detailed comparison of the results
from the different pipelines is ongoing.

The tight correlation in Figure~\ref{fig:01} between \Dnu\ and \numax\ was
discussed by \citet{HKB2009} and \citet{SCB2009}, and is seen here to
extend to red giants with much higher \numax.  This and other correlations
between global oscillation parameters, and their connection to fundamental
stellar parameters, will be discussed in future papers.

\section{Results}

\subsection{Frequency analysis}
\label{sec:freqs}

We now consider a subset of 50 of the low-luminosity red giants for more
detailed analysis, chosen as having the best signal-to-noise ratios.  These
stars are indicated in Figure~\ref{fig:01} with filled symbols.
Figure~\ref{fig02} (left panel) shows power spectra for six of them,
spanning the full range of \numax\ being considered.  Each power spectrum
shows a regular series of peaks, which is the clear signature of solar-like
oscillations.

Solar-like oscillations are p modes with high order and low degree, and the
observed frequencies in main-sequence stars are reasonably well-described
by the following relation:
\begin{equation}
  \nu_{n,l} \approx \Dnu (n + \half l + \epsilon) - l(l+1) D_0.
        \label{eq:asymptotic}
\end{equation}
Here, $n$ (the radial order) and $l$ (the angular degree) are integers.
The form of Equation~\ref{eq:asymptotic} is motivated by theoretical
calculations: for relatively unevolved stars, an asymptotic expansion
\citep{Tas80,Gou86} shows that $\Dnu$ is approximately the inverse of the
sound travel time across the star, while $D_0$ is sensitive to the sound
speed gradient near the core and $\epsilon$ is sensitive to the surface
layers.

We conventionally define three small frequency separations: $\dnu{02}$ is
the spacing between adjacent modes with $l=0$ and $l=2$, $\dnu{13}$ is the
spacing between adjacent modes with $l=1$ and $l=3$, and $\dnu{01}$ is the
amount by which $l=1$ modes are offset from the midpoint between the $l=0$
modes on either side.  If Equation~\ref{eq:asymptotic} holds then it
follows that $\dnu{02} = 6 D_0$, $\dnu{13} = 10 D_0$ and $\dnu{01} = 2
D_0$.  {We shall use Equation~\ref{eq:asymptotic} as a guide to the
analysis of the observed frequencies, although its physical interpretation
may be open to question in the case of red giants.}

{To determine oscillation frequencies for each of the 50 stars in our
subset, we needed to locate the highest peaks in the power spectra.}  We
did this using conventional iterative sine-wave fitting, sometimes called
pre-whitening or CLEAN\@.  This involves finding the highest peak in the
power spectrum, subtracting the corresponding sinusoidal variation from the
time series and then re-calculating the power spectrum in an iterative
process.  We retained only those peaks that lay in the frequency range
\numax $\pm$ 5\Dnu\ and had heights greater than 4.5 times the fitted
background level (in amplitude).

Damping of solar-like oscillations causes the observed power from each mode
to be spread into multiple peaks under a Lorentzian envelope.  If the
length of the observations is significantly greater than the typical mode
lifetime, one therefore expects several peaks to be extracted for each
mode.  In fact, the modes are only barely resolved (see
Section~\ref{sec:collapsedechelle}) and so we expect iterative sine-wave
fitting to give good results.

\subsection{Semi-scaled \'Echelle Diagram}
\label{sec:semiechelle}

A powerful method to study the mode frequencies of solar-like oscillations
is to plot them modulo the large separation, in a so-called \'echelle
diagram \citep{GFP83}.  We have displayed the frequencies of all 50 stars
in a single \'echelle diagram by using the scaling technique described by
\citet{B+K2010}, in an effort to carry out ``ensemble asteroseismology''.
The result is shown in Figure~\ref{fig03}{\it a}.  To make this diagram, we
first divided the frequencies of each star by \Dnu\ (that is, scaling them
to have a large separation of unity).  We used the \'echelle diagram to
identify the radial modes and then fine-tuned the scaling factor (\Dnu) by
up to a few percent to align them on a single line in the \'echelle
diagram, shown as a solid line in Figure~\ref{fig03}{\it a}.  As discussed
by \citet[][their Method~2]{B+K2010}, this equates to assigning all stars
the same value of~$\epsilon$.  {The fact that all the stars can be aligned
indicates this is a valid approximation.}  This alignment is also
clearly seen in the right-hand panel of Figure~\ref{fig02}, which shows
power spectra plotted against scaled frequencies (i.e., divided by \Dnu).

The \'echelle diagram in Figure~\ref{fig03}{\it a} differs from those shown
by \citet{B+K2010} in one important respect: only the horizontal axis has been
scaled.  In the vertical direction, we have plotted the original
frequencies without scaling, which spreads out the points and allows us to
look for variations with~\numax.  Such a plot might be called a semi-scaled
\'echelle diagram.

\subsection{Collapsed \'Echelle Diagram and Mode Lifetimes}
\label{sec:collapsedechelle}

We can collapse the \'echelle diagram in the vertical direction in order to
define the ridges more clearly.  Figure~\ref{fig03}{\it b} shows this using
a simple histogram of the peaks in Figure~\ref{fig03}{\it a}, and the
ridges are clearly visible.  Rather than counting the number of peaks above
a threshold, another approach is to sum the total power.  To do this, we
restricted each background-corrected power spectrum to the same frequency
range as before ($\numax \pm 5\Dnu$), then divided the frequency scale by
the large separation (using the same fine-tuned value as above) and folded
the spectrum modulo unit spacing.  These folded spectra were summed for all
50 stars and then smoothed slightly (with a boxcar of width 0.01\Dnu) to
give the result shown in Figure~\ref{fig03}{\it c}.

The clarity of the ridges in Figure~\ref{fig03} allows us to make some
statements about mode lifetimes.  As mentioned in Section~\ref{sec:freqs},
damping causes each mode
to be spread into multiple peaks under a Lorentzian envelope with a
full-width at half maximum (FWHM) of $1/(\pi\tau)$, where $\tau$ is the
mode lifetime.  There has been considerable discussion of the lifetimes of
red-giant oscillations, particularly regarding whether the modes are
sufficiently long-lived to allow their frequencies to be measured with
sufficient accuracy to be useful for asteroseismology
\citep{H+G2002,SKB2006,BMDeR2007,DBS2009}.  The \corot\ results have
clearly dismissed this worry for red clump stars \citep{DeRBB2009,CDRB2010}
and our \Kepler\ data do the same for the low-luminosity stars.  This is
excellent news for asteroseismology.

Figure~\ref{fig03} was constructed by aligning the radial modes on the
vertical solid line.  However, the resulting width of the $l=0$ ridge is
not zero, which arises from at least three factors: {(1)~the resolution
limit set by the duration of the observations (34\,d), which leads to peaks
in the power spectrum having FWHM 0.3\,\muHz;} (2)~variations in \Dnu\
with frequency in individual stars over the range being considered
(curvature in the \'echelle diagram); and (3)~broadening of power due to
finite mode lifetimes (see above).  The observed narrowness of the $l=0$
ridge 
sets upper limits on the second and third of these effects.  Here, as
discussed above, we are particularly interested in the mode lifetimes.

The FWHM of the $l=0$ peak in Figure~\ref{fig03}{\it c}, found by
subtracting the background and fitting a Lorentzian, is 0.033\Dnu.  This
sets an upper limit on the intrinsic linewidths of these modes.  However,
so far our measurement is in terms of \Dnu, which varies over the sample.
The median \Dnu\ is 11.5\,\muHz, in which case the measured width equates
to a lower limit on the mode lifetimes of about 10\,days.  To investigate
further, we divided the sample into four bins according to \Dnu\ and formed
the collapsed \'echelle diagram for each subset, as before.  We found the
widths of the $l=0$ ridges in all four cases to be about the same in
absolute terms ($\sim$0.4\,\muHz), which is only slightly greater than the
FWHM set by the resolution of the data (see above).  We conclude that the
mode lifetimes in these red giants is at least 10\,days, possibly much
greater.

\subsection{Small Separations and the C-D Diagram}

The $l=2$ ridge is closely parallel to the $l=0$ ridge, implying that
\dnu{02} is a nearly constant fraction of \Dnu\ for the stars in our
sample, with $\dnu{02} \approx 0.125\Dnu$ (Figure~\ref{fig03}{\it c}).  This
frequency scaling indicates that, to a good approximation, the stars are
homologous.

We were able to measure the mean small separation \dnu{02}\ for 38 stars in
our sample.  Figure~\ref{fig04}{\it a} shows the results in a so-called C-D
diagram \citep{ChD88b}, which plots \dnu{02} versus \Dnu.  The dashed line
in Figure~\ref{fig04}{\it a} shows a linear fit, which has parameters
\begin{equation}
  \dnu{02} = (0.122 \pm 0.006)\Dnu + (0.05 \pm 0.08)\,\muHz.
 \label{eq:CD}
\end{equation}
The spread of points about this relation presumably reflects a spread in
stellar masses, and is better seen by plotting the ratio of the two
separations, as shown in Figure~\ref{fig04}{\it b}
\citep[see][]{R+V2003,FChDT2005,Maz2005}.

The dotted line in Figure~\ref{fig03} marks the midpoint between $l=0$
modes.  Interestingly, the $l=1$ ridge is centered slightly to the right of
this line.  This indicates that the average small separation \dnu{01} in
these red giants is negative, whereas it is positive in the Sun and the
handful of other main-sequence stars for which it has been measured:
$\alpha$~Cen~A \citep{BK04}, $\alpha$~Cen~B \citep{KB05}, $\gamma$~Pav
\citep{MDM2008}, and $\tau$~Cet \citep{teixeira09}.  A negative value for
\dnu{01} also indicates that Equation~\ref{eq:asymptotic} is not satisfied
in detail for these red giants, a point already noted for the G8 giant
HR~7349 by \citet{CDRB2010}, based on \corot\ observations.

\subsection{Mixed Modes with $l=1$}
\label{sec:mixedmodes}

\begin{figure*}
\epsscale{1.0}
\plotone{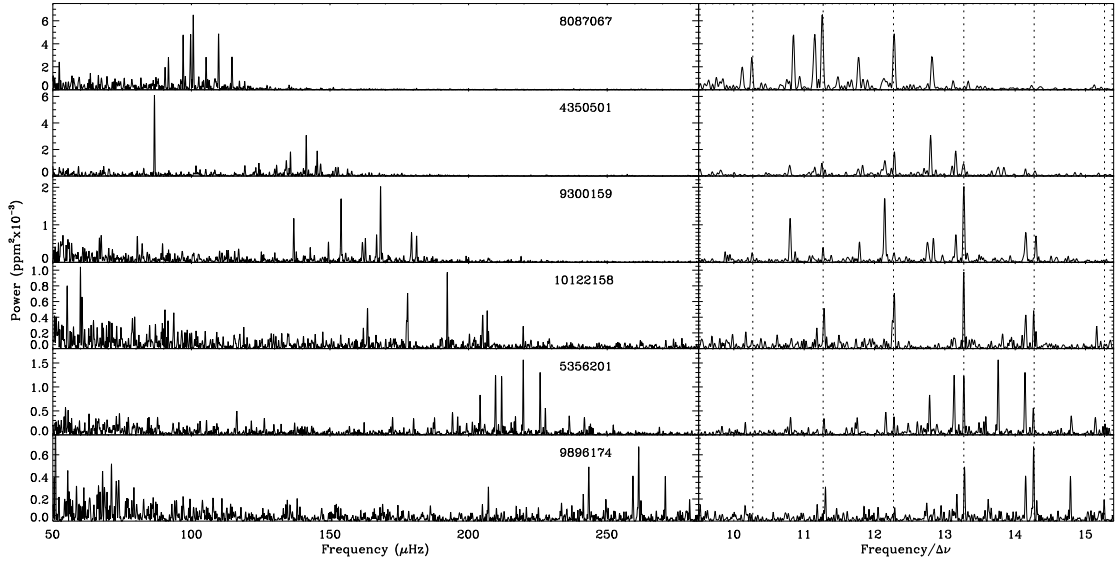}
\caption{{\it Left:} Power spectra of six representative low-luminosity red
giants.  {\it Right:} The same power spectra plotted against scaled
frequency (see Section~\ref{sec:semiechelle}).  \new{The dotted lines are
equally spaced, having unit separation and being aligned with the $l=0$
modes.}  Stars are labelled with identification numbers from the {\em
Kepler Input Catalog} (KIC; \citealt{LBM2005}). }
\label{fig02}
\end{figure*}

\begin{figure}
\epsscale{0.5}
\plotone{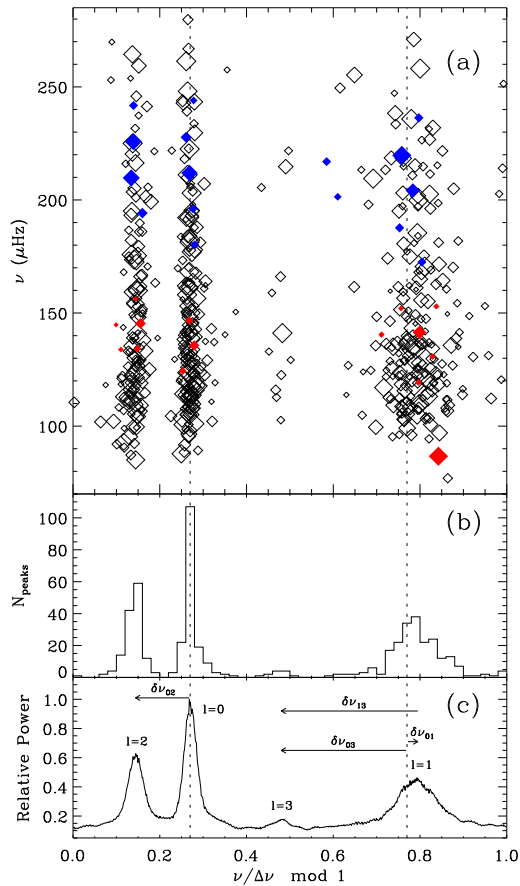}
\caption{({\it a}) Semi-scaled \'echelle diagram for the sample of 50
  low-luminosity red giants (see Section~\ref{sec:semiechelle}).  Symbols
  indicate extracted frequencies, with symbol sizes proportional to
  amplitude.  Filled blue and red symbols are frequencies of the stars
  KIC~5356201 and KIC~4350501, respectively (see Figure~\ref{fig02} and
  Section~\ref{sec:mixedmodes}).  The value of \Dnu\ for each star was
  fine-tuned in order to align the $l=0$ modes on the left-most vertical
  dotted line.  The other dotted line \new{lies} exactly 0.5 away, marking
  the midpoint between $l=0$ modes.  ({\it b}) Histogram of the points
  shown in panel~{\it a}.  ({\it c}) Folded and scaled power spectra,
  collapsed over all stars in the sample and smoothed slightly (see
  Section~\ref{sec:semiechelle}).  The arrows show the definitions of the
  small separations \dnu{01}, \dnu{02}, \dnu{03}, \new{and \dnu{13},} with
  the sign convention that left-pointing arrows correspond to positive
  separations (see Section~\ref{sec:freqs}).  }
\label{fig03}
\end{figure}

\begin{figure}
\epsscale{0.55}
\plotone{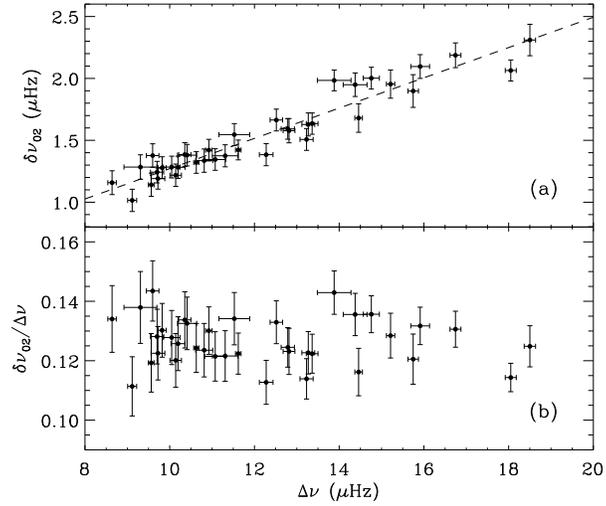}
\caption{({\it a}) The so-called C-D diagram \citep{ChD88b} of \dnu{02}
versus \Dnu\ for the 38 stars in our sample for which both quantities were
reliably determined.  The dashed line shows a linear fit. ({\it b}) The
ratio \dnu{02}/\Dnu\ versus \Dnu\ for the same stars, to better show the
scatter about the trend. }
\label{fig04}
\end{figure}

The $l=1$ ridge in Figure~\ref{fig03} is notably broader than the others.
{We do not attribute this to shorter mode lifetimes, but rather to the
presence of mixed modes whose frequencies are shifted by avoided crossings.
Mixed modes occur in evolved stars and have the characteristics of p modes
in the envelope of the star and of g modes in the interior
\citep[e.g.,][]{ASW77}.}  A typical example is the star KIC~5356201 (see
Figure~\ref{fig02} and filled blue symbols in Figure~\ref{fig03}{\it a}).  The
$l=0$ and 2 modes are quite close to their respective ridges, whereas there
is much more scatter about $l=1$.  A more extreme example is the star
KIC~4350501 (see Figure~\ref{fig02} and filled red symbols in
Figure~\ref{fig03}{\it a}), which shows a particularly high peak at
87\,\muHz, well below \numax.  The proximity of this peak to the $l=1$
ridge indicates that it is likely to be a mixed mode.  \new{A similar
situation is seen in the F5 subgiant Procyon, for which ground-based
velocity observations show a narrow peak well below \numax\ that lies close
to the $l=1$ ridge in the \'echelle diagram \citep{BKC2010}.}  The large
peak height can be explained by the fact that mixed modes are expected to
have longer lifetimes (smaller linewidths) than pure p modes because they
have larger mode inertias \citep[e.g.,][]{ChD2004}.

\begin{figure}
\epsscale{0.55}
\plotone{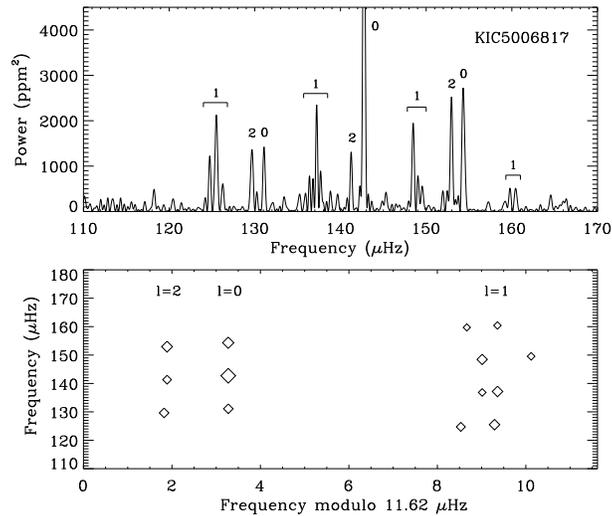}
\caption{Power spectrum and \'echelle diagram for the star KIC~5006817,
  illustrating multiple $l=1$ peaks per order.  Note that the highest peak
  in the power spectrum extends beyond the plot limits (to 8400\,ppm$^2$).
  }
\label{fig:nice}
\end{figure}

Another good example is KIC~5006817, shown in Figure~\ref{fig:nice}, where
we clearly see multiple $l=1$ peaks in each order \citep[see also
KIC~9904059 = KOI~145 in Figure~1 of][]{GBChD2010}.  Each of these
clusters of peaks is reminiscent of a single mode that is heavily damped
(see Section~\ref{sec:collapsedechelle}).  However, as mentioned above, mixed
modes are expected to have longer lifetimes than pure p modes and hence to
produce narrower -- not broader -- peaks.  Indeed, this phenomenon of
multiple $l=1$ peaks is probably what led previous authors to report short
mode lifetimes in red giants (see Section~\ref{sec:collapsedechelle}).  Guided
by the theoretical work of \citet{DBS2009}, we instead interpret the
observed properties of the $l=1$ ridge in terms of mixed modes.

\citet{DBS2009} discussed theoretical amplitudes of solar-like oscillations
for a range of models on the red giant branch.  Their models show mostly
regular frequency patterns for evolved giants (\numax\ $<$ 80\,\muHz), but
predict more complex spectra for low-luminosity giants.  This is mainly
explained by the less efficient radiative damping in the cores of these
stars, leading to a stronger interaction of p-mode and g-mode cavities, and
by the less efficient trapping due to the smaller density contrast between
the core and the envelope.  As a consequence, more non-radial modes could
be observed than only those trapped in the envelope.  This is particularly
expected to affect $l = 1$ modes because the evanescent region just above
the H-burning shell is smaller for these modes \citep[see
also][]{DGH2001,ChD2004}.  Our observation of a broadened $l=1$ ridge with
a small cluster of modes per order supports the theoretical predictions by
\citet{DBS2009}.  Observations of frequency spectra in a sample of higher
luminosity red giants would be a desirable follow-up study.

\subsection{Detection of $l=3$ Modes}
\label{sec:leq3}

In addition to ridges corresponding to $l=0$, 1 and 2, we see clear
evidence in Figure~\ref{fig03} for modes with $l=3$.  The detection of
these modes with photometry is difficult because of the very low amplitudes
that result from geometric cancellation.  The only previous report of
solar-like oscillations with $l=3$ from photometry (except for the Sun) is
a probable detection using \corot\ in the G0 main-sequence star HD~49385
\citep{DBM2010}.  The ability to detect $l=3$ modes from \Kepler\
observations of red giants is a significant bonus.

In Figures~\ref{fig03}{\it a} and~{\it b} there are 12 peaks that fall
along the ridge that we identify with $l=3$, and these arise from 10 stars
in the sample of 50.  These are presumably peaks that, due to the
stochastic nature of the excitation, happen to rise above our detection
threshold in these observations.  A more objective approach is to sum the
power from all stars, as shown in Figure~\ref{fig03}{\it c}, and we again
see clear evidence for the $l=3$ ridge.

How does the position of the $l=3$ ridge compare with expectations?  As
mentioned in Section~\ref{sec:freqs}, it is conventional to measure $l=3$
relative to $l=1$, via the small separation~\dnu{13}.  However, given that
the position of $l=1$ for these red giants is anomalous, it seems sensible
to instead measure $l=3$ relative to $l=0$.  We therefore suggest using a
new small separation, \dnu{03}, defined (by analogy to \dnu{01}) as the
amount by which the $l=3$ modes are offset from the midpoint between the
$l=0$ modes on either side.  If follows that $\dnu{03} = \dnu{01} +
\dnu{13}$.  From Figure~\ref{fig03}{\it c} we see that $\dnu{03} \approx
0.28\Dnu$.  If the asymptotic relation held (Equation~\ref{eq:asymptotic}), we
would expect $\dnu{03} = 12D_0$.  In that case the ratio between \dnu{03}
and \dnu{02} would be 2.0, whereas we observe a ratio of 2.2.  These
results should provide valuable tests for theoretical models.


\section{Conclusions}

These results, based on the analysis of only 34 days of data, show the
tremendous potential of \Kepler\ long-cadence data for the study of
solar-like oscillations in red giants.  As more data become available, the
longer time series will allow detailed studies of stars covering the whole
range of evolutionary states along the red giant branch and provide
stringent observational tests for theories of stellar structure and
evolution in this part of the H-R diagram.

\acknowledgments We gratefully acknowledge the entire \Kepler\ team, whose
  outstanding efforts have made these results possible.  Funding for this
  Discovery mission is provided by NASA's Science Mission.  
  Grants were received from the European Research Council  (ERC/FP7/2007--2013 n$^\circ $227224,
  PROSPERITY), and K.U.Leuven (GOA/2008/04).

{\it Facilities:} \facility{Kepler}

\end{document}